\documentclass[%
 reprint,
 amsmath,amssymb,
 aps,
]{revtex4-2}

\usepackage{graphicx}
\usepackage{dcolumn}
\usepackage{bm}
\usepackage{amsmath,amsfonts,amssymb}
\usepackage{subfig}
\usepackage{tikz}

\begin{document}

\preprint{APS/123-QED}

\title{A Practitioner’s Guide to Quantum Algorithms for
Optimisation Problems}

\author{Benjamin C. B. Symons}
\affiliation{%
 The Hartree Centre \\ STFC, Sci-Tech Daresbury \\ Warrington WA4 4AD \\ United Kingdom
}%
\author{David Galvin}%
\affiliation{%
 The Hartree Centre \\ STFC, Sci-Tech Daresbury \\ Warrington WA4 4AD \\ United Kingdom
}%
\author{Emre Sahin}
\affiliation{%
 The Hartree Centre \\ STFC, Sci-Tech Daresbury \\ Warrington WA4 4AD \\ United Kingdom
}%
\author{Vassil Alexandrov}
\affiliation{%
 The Hartree Centre \\ STFC, Sci-Tech Daresbury \\ Warrington WA4 4AD \\ United Kingdom
}%
\author{Stefano Mensa}
 \email{stefano.mensa@stfc.ac.uk}
\affiliation{%
 The Hartree Centre \\ STFC, Sci-Tech Daresbury \\ Warrington WA4 4AD \\ United Kingdom
}%

\date{\today}

\begin{abstract}
Quantum computing is gaining popularity across a wide range of scientific disciplines due to its potential to solve long-standing computational problems that are considered intractable with classical computers. One promising area where quantum computing has potential is in the speed-up of \textit{NP}-hard optimisation problems that are common in industrial areas such as logistics and finance. Newcomers to the field of quantum computing who are interested in using this technology to solve optimisation problems do not have an easily accessible source of information on the current capabilities of quantum computers and algorithms. This paper aims to provide a comprehensive overview of the theory of quantum optimisation techniques and their practical application, focusing on their near-term potential for noisy intermediate scale quantum devices.
The paper starts by drawing parallels between classical and quantum optimisation problems, highlighting their conceptual similarities and differences. Two main paradigms for quantum hardware are then discussed: quantum annealing and gate-based quantum computing. While quantum annealers are effective for some optimisation problems, they have limitations and cannot be used for universal quantum computation. In contrast, gate-based quantum computers offer the potential for universal quantum computation, but they face challenges with hardware limitations and accurate gate implementation. The paper provides a detailed mathematical discussion with references to key works in the field, as well as a more practical discussion with relevant examples.
The most popular techniques for quantum optimisation on gate-based quantum computers, the quantum approximate optimisation (QAO) algorithm and the quantum alternating operator ansatz (QAOA) framework, are discussed in detail. However, it is still unclear whether these techniques will yield quantum advantage, even with advancements in hardware and noise reduction. The paper concludes with a discussion of the challenges facing quantum optimisation techniques and the need for further research and development to identify new, effective methods for achieving quantum advantage.
\end{abstract}

\maketitle


\section{\label{sec:level1}Introduction}

The principles of quantum mechanics have enabled the development of a new paradigm of computing, known as quantum computing. This innovative approach leverages the principles of superposition and entanglement to offer faster and more efficient solutions to specific industrial challenges that classical computers struggle to address.  For example, the ability to find prime factors of large numbers very quickly on a quantum computer has the potential to break widely used encryption algorithms that rely on the difficulty of this problem \cite{ShorAlgo}. Additionally, quantum computing has implications for drug discovery \cite{Mensa_2023}, materials science \cite{Bauer}, and machine learning algorithms \cite{Biamonte}, amongst other fields. While the aforementioned applications of quantum computing are promising, a significant amount of research is being carried out on the development of quantum-enhanced optimisation techniques for logistics \cite{vogiatzis2013combinatorial, NSW, clark2019towards}, automotive \cite{yarkoni2020quantum}, finance \cite{Slate2021quantumwalkbased,phillipson2021portfolio} and the defence sector \cite{krelina2021quantum}. \\

In the classical domain, optimisation problems \cite{boyd2004convex} are defined as mathematical problems that involve finding the best solution from a set of possible solutions. The goal of optimisation is to find the solution that maximizes or minimizes an objective function - a measure of success - subject to a set of constraints limiting the allowable solutions. Typically, classical optimisation algorithms rely on searching through the space of all possible solutions, evaluating each solution until the best one is found. This process can be time-consuming and inefficient, especially for problems with many variables or constraints. As such the search typically proceeds according to some heuristic approach that can speed up the process considerably. A well-known example is the Traveling Salesman Problem (TSP) \cite{applegate2011traveling}, in which a salesman must visit a set of cities, each only once, and return to his starting point while minimizing the total distance travelled. This is a combinatorial optimisation problem, which means that the number of possible solutions grows combinatorially with the number of cities. As a result, finding the optimal solution to the TSP for many cities becomes time-consuming and computationally expensive for classical computers. In fact, the TSP is an example of an \textit{NP}-hard problem, which means that there is no known efficient classical algorithm that can solve it in polynomial time. The best classical algorithms for the TSP rely on heuristics and approximation methods, which can still take a long time to reach a solution, especially for large instances of the problem.\\

Quantum computers are, in principle, inherently well suited to the task of solving optimisation problems thanks to the key phenomena of superposition and entanglement \cite{nielsen2002quantum}. Superposition refers to the ability of qubits – the basic building blocks of quantum computers - to exist in multiple states simultaneously. It is commonly claimed that superposition will enable a quantum computer to explore all possible solutions to a problem simultaneously, thereby leading to an advantage over classical computers. Such claims must be considered carefully because, in reality the probabilistic nature of measurement outcomes in quantum mechanics means that superposition alone is not enough to yield an advantage. Superposition must be combined with clever algorithm design in order for quantum computers to have high probabilities of outputting useful information. Entanglement, on the other hand, refers to the strong correlations that exist between qubits in a quantum system. A key property of this correlation is that it is non-local i.e. two qubits can be entangled even if they are separated by arbitrarily large distances. Entanglement is a fundamentally non-classical phenomenon and it is believed that this will be a key resource that enables quantum advantage. Indeed, entanglement is a key ingredient in many quantum optimisation algorithms, such as quantum annealing and variational quantum algorithms (discussed in this review), which are specifically designed to exploit the benefits of quantum mechanics for solving optimisation problems. However, the actual mechanisms by which we can best exploit entanglement in algorithms are not currently known. This is primarily because entanglement is still not well understood in these contexts. Finally, another key reason optimisation has been an early application area of quantum computing is the existence of mappings of many problem instances onto Ising problems which in turn can be mapped onto a quantum computer. In general, it is highly non-trivial to map a problem to a form that can be executed on a quantum computer. As such the existence of a simple mapping is itself a good reason to examine a class of problems. \\

One of the most exciting aspects of quantum computing is the concept of quantum advantage \cite{daley2022practical}. Although exact definitions differ somewhat, it is generally accepted that quantum advantage will have been achieved once a quantum computer can solve a practically useful problem (significantly) faster than any classical computer. The potential advantages that quantum computers offer in solving complex optimisation problems have made the topic highly popular among both the scientific community and the public. This is evident from the significant media coverage that it has received. However, there are several misconceptions surrounding this topic, particularly regarding the speed and universality of the optimisation problems that can be addressed. While it is correct that quantum computers may have the capacity to solve some problems faster than classical computers, true quantum advantage has not yet been established in practice. Additionally, an optimisation problem that can genuinely benefit from a quantum device has yet to be identified.\\

Approaching quantum optimisation to solve real-world industrial challenges can be a daunting task for practitioners exploring quantum computing as a potential solution. Distinguishing between facts and myths among the plethora of available articles on this subject can be challenging. Furthermore, there is currently no clear understanding of the potential benefits of algorithms or any prospective quantum advantage. In this work, we aim to help practitioners from various disciplines involving optimisation problems to understand the state of the art of using quantum computers and algorithms to solve optimisation problems. Specifically, we provide a parallelism between classical and quantum optimisation problems, analyzing analogies and differences, and explaining the kind of problems that will, in principle, benefit from quantum acceleration. We analyze the difference between digital and analog quantum computers and why quantum algorithms will perform differently. All of this information lays the foundation to understand what a quantum computer can and cannot do to solve optimisation problems efficiently, with the ultimate benefit of setting clear community expectations. For interested readers, we provide a detailed mathematical overview of the most common quantum algorithms for optimisation, such as QAOA, useful for understanding why such algorithms may lead to quantum advantage. Each section provides an overview of the some of the most relevant work in the field, redirecting the reader towards compelling practical case studies to inspire new work. Finally, we conclude this work by discussing the pros and cons of applied quantum computing to optimisation problems, as well as future directions.

\section{Classical Optimisation}

The field of classical optimisation encompasses a large number of specific problems that often have useful, practical applications across a wide variety of fields. Depending on whether the variables are discrete or continuous, optimisation problems can be divided in two classes, \textit{discrete optimisation} and \textit{continuous optimisation}. In this work we focus almost exclusively on discrete optimisation problems. In this section we want to provide the reader with a general overview of the most common optimisation problems encountered by the scientific community, explaining why these can be computationally intractable using classical computing methodologies

\subsection{\label{sec:level2}Discrete Optimisation}

Discrete optimisation problems - also known as \textit{combinatorial} optimisation problems \cite{korte2011combinatorial,Korte2018-yg} - involve decision variables that take on discrete or categorical values, such as integers, binary values, or categories. Combinatorial optimisation problems involve searching through a finite or countable set of possible solutions, making this class of problems computationally intractable, as the number of possible solutions can be very large. Many combinatorial optimisation problems are known to be \textit{NP}-hard problems, meaning that as the size of the problem increases, the time required to find an optimal solution grows exponentially as well (assuming P$\neq$\textit{NP}). This makes this class of optimisation problems incredibly challenging to be solved exactly in a satisfactory amount of time. There has been a great deal of effort from the community directed towards solving these problems as accurately and as quickly as possible. This has resulted in a plethora of heuristic algorithms. Rather than exhaustively searching the space of all possible solutions for the exact solution, heuristic methods are designed to find a good-quality approximate solution. These methods are typically able to find solutions efficiently (in polynomial time) but, this comes at the cost of providing a sub-optimal solution \cite{papadimitriou1998combinatorial,blum2003metaheuristics}. \\

In practice, the choice of algorithm depends on the size and structure of the problem instance, as well as the desired trade-off between solution quality and computation time. For small problem instances, exact methods may be preferred, as they can guarantee optimality. However, for large problem instances, exact methods may become computationally infeasible, and approximate methods may be necessary to find a good-quality solution within a reasonable amount of time. Many combinatorial optimisation problems are in fact known to be \textit{APX}-hard. The \textit{APX} complexity class is comprised of \textit{NP} problems that are able to approximate in polynomial time to within a constant multiplicative factor of the optimal solution. This has allowed for the development of many very useful approximate classical algorithms for a wide variety of combinatorial optimisation problems. It is generally believed that quantum computers will not be able to exactly solve \textit{NP}-hard optimisation problems efficiently. This is because the quantum versions of these problems are \textit{QMA}-hard\cite{qma-hard} (the \textit{QMA} complexity class can be thought as the quantum analogue of \textit{NP}). Quantum algorithms are therefore typically approximate in nature and the real question is whether or not quantum computers are able to yield better and/or faster approximate solutions relative to classical algorithms. \\

In this subsection, we will discuss some of the most important classes of discrete combinatorial optimisation problems, including the travelling salesman problem, Max-Cut, Max-Flows and the Knapsack problem.

\subsubsection{The Travelling Salesman Problem}

The travelling salesman problem (TSP) is perhaps the most well-known combinatorial optimisation problem. In order to setup the problem, consider a salesman that must visit a number of cities, each with various distances between them. The canonical form of the problem is to find the shortest path that visits all cities. The problem lends itself well to being mapped to a graph problem. Given an undirected, weighted graph $G = (V, E)$, each city is a vertex $V$, and cities are connected by edges $E$ with weights representing distances. Despite the simple setup of the problem it is very challenging to solve, in fact it is \textit{NP}-hard to solve exactly. There are a large number of applications of TSP (or variants of the problem that have extra constraints) including logistics, circuit board manufacturing and planning to name but a few.

\subsection{\label{sec:citeref}The Max-Cut Problem}

\begin{figure}
\centering
\begin{tikzpicture}
    \node[shape=circle,draw=black, fill=black, thick] at (0,0) {};
    \node[shape=circle,draw=black, thick] at (0,2) {};
    \node[shape=circle,draw=black, thick] at (2,0) {};
    \node[shape=circle,draw=black, fill=black, thick] at (2,2) {};
    \node[shape=circle,draw=black, fill=black, thick] at (1.0,1.0) {};

    \draw[-, thick](0,0) -- (0,2) [right];
    \draw[-, thick](0,0) -- (2,0) [right];
    \draw[-, thick](2,0) -- (2,2) [right];
    \draw[-, thick](0,2) -- (2,2) [right];
    \draw[-, thick](1,1) -- (0,0) [right];
    \draw[-, thick](1,1) -- (2,0) [right];
    \draw[-, thick](1,1) -- (0,2) [right];
    \draw[-, thick](1,1) -- (2,2) [right];
\end{tikzpicture}
\caption{An illustration of a simple 5 node Max-Cut problem. The coloring of the nodes (black and white) demonstrates one example of a partitioning into two separate sets that yields a maximum cut.}
\label{max-cut-fig}
\end{figure}
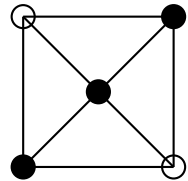

The maximum cut problem is a well-known optimisation problem that requires partitioning the vertices of an undirected graph into two disjoint sets, such that the number of edges between the two sets is maximized. Given a graph $G = (V,E)$, a cut represents a partition of the graph into two distinct subgraphs such that no vertex is shared between either graph. The Max-Cut problem consists of finding a cut in the graph such that the number of edges between the graph partitions is maximised. More generally in the case of weighted graphs, the problem extends to finding a cut that maximises the cumulative value of the edges. An example of the Max-Cut problem is shown in Figure \ref{max-cut-fig}.\\

The Max-Cut problem has applications in a variety of fields, including computer vision, social network analysis, and computational biology. It is known to be \textit{NP}-hard, which means that finding the optimal solution is generally believed to be computationally intractable for large instances of the problem. Therefore, many heuristic (approximate) algorithms have been proposed to solve the problem efficiently. The Max-Cut problem can be mapped to an Ising problem which makes it a good candidate for a quantum computer.

\subsubsection{The Max-Flow problem}

A natural problem in transport and logistics is to attempt to model and optimise the flow of goods and services to customers. In the most basic example, one could consider the transport of goods from a depot to a destination along a network where certain routes have only limited capacities for transport. A problem of this form can be mapped to a graph problem. The goal is to maximise the flow through the graph given the constraints. A better value for the max flow corresponds to a more efficient delivery process. \\

Given a graph $G(V,E)$ with edge capacities $c_{ij}$, a flow $f$ is defined as a function $f: E \rightarrow \mathbb{R}^+$ such that $f(i,j) \leq c_{ij}$. Given two vertices $s$ and $t$ labelled the source and sink respectively, we constrain the flow such that it is equal when entering and exiting at all vertices except the source and sink vertices, 

\begin{equation}
    \sum_{u:(u,v) \in E}f(u,v) = \sum_{u:(v,u)\in E}f(v,u) \ \ \ \forall v \in V\backslash \{s,t\}.
\label{max-flow-1}
\end{equation}

A natural consequence of this constraint is that flow is conserved. For a single source-sink pair with such a constraint, the flow may be given a value $C_f$ describing the net output of the source vertex $s$, that is,

\begin{equation}
    C_{f} := \sum_{u:(s,u)\in E} f(s,u) - \sum_{u:(u,s) \in E} f(u,s).
\end{equation}

The maximum flow problem concerns finding a flow $f^*$ such that $C_{f^*}$ is maximum, $C_{f^*} \geq C_{f} \ \ \forall f:E \rightarrow \mathbb{R}^+$. The maximum flow problem represents a canonical optimisation problem over a weighted graph. Real world applications often require additional or different constraints on transport. For example, the Vehicle Routing Problem \cite{toth2002vehicle} which requires that multiple vehicles service multiple destinations (or have different source depots). Variations to the problem include the introduction of capacity constraints on the vehicles (CVRP) or time windows for delivery (VRPTW).

\subsubsection{The Knapsack Problem}

The knapsack problem is an \textit{NP}-complete combinatorial optimisation problem that has applications in resource allocation. The knapsack problem is usually stated as follows: given a set of items, each with a given weight and value, determine the a collection of items that has total weight less than some maximum whilst also having maximum possible value. There are several variations on the problem, the simplest is the so-called `0-1 knapsack problem' that restricts each item to a single copy i.e. each item has $x_i \in \{0,1\}$ copies. This problem is stated mathematically as,

\begin{equation}
    \text{Maximise} \sum_i v_i x_i,
\end{equation}

\begin{equation}
    \text{subject to} \sum_i w_i x_i \leq W,
\end{equation}

where $v_i$ and $w_i$ are values and weights respectively and $W$ is the maximum total weight. The problem can be extended such that multiple copies of each item are permitted or also to include multiple weights e.g. monetary cost and mass of each item.

\subsection{Continuous Optimisation}

Continuous optimisation problems involve variables that take on continuous values, such as real numbers \cite{strang1986introduction}. The objective function in a continuous optimisation problem maps a set of decision variables to a scalar value that represents the objective of the problem. The decision variables can take on any value within a specified range or domain. The goal is to find the values of the decision variables that minimize or maximize the objective function. The constraints can be expressed as equations or inequalities, and they limit the feasible region of the decision variables. Examples of continuous optimisation problems include finding the shortest path between two points in a curved space, determining the optimal allocation of resources for a production process, or optimising the design of a complex system. Despite the fact that the fields of continuous and discrete optimisation both aim to solve optimisation problems, they are in fact rather different and separate fields of study. In the remainder of this work we focus on discrete optimisation as this has been the sub-field of optimisation that has been studied the most extensively in the context of quantum computing.

\section{Optimisation on Analog Quantum Computers}

Analog computing is perhaps an unfamiliar concept in the age of digital computing. Most are familiar with the digital (gate-based) model that utilises a series of discrete operations to perform a computation. Analog computing on the other hand, involves the continuous time evolution of an initial state to a final state. The goal is to setup and evolve the analog computer in such a way that the final state is useful e.g. it is the solution to a problem. There is a conceptual shift that has to be made when transitioning from thinking about digital to analog computing. In the digital model of computation it is possible to understand computation without considering the underlying hardware on which the computation is implemented. In the analog model this is no longer possible because the nature of the continuous evolution of the system is very much tied to the nature of the system itself i.e. the hardware. It is not possible to formulate a general hardware-independent model of analog computation.\\

In the case of quantum computing, the boundary between analog and digital is not as sharp as it is in the classical realm. Despite the fact that, in the gate-based model, computation is thought of as a series of discrete gates acting on a quantum state, in reality there is still a continuous time evolution of a quantum system. This means that, in some sense, all quantum computers are analog. In both analog and digital quantum computers, the quantum state prior to measurement is some continuous state that exists in the Hilbert space of the quantum computer. However, after measurement both types of quantum computer output a bit string (a binary number) i.e. both analog and digital quantum computers produce a digital output. Despite these similarities, one of the main distinguishing features of digital quantum computers is universality. We will discuss the concept of universality in more detail in the section on digital quantum computers but, it suffices to say here that analog quantum computers are not, in general, universal. This means that the analog quantum computers that exist today are only able to tackle very specific problems. The kind of problem a given analog quantum computer is able to solve is closely related to the form of the hardware itself. \\

At the time of writing, D-Wave has the most mature technology (superconducting qubits) in the space of analog quantum computing, specifically they develop quantum annealers \cite{D-wave,D-wave2023}. Recently, there have been considerable advances in neutral atom technology, a different approach that is able to blend the more traditional model of analog quantum computation with the gate-based model. This allows companies such as Pasqal \cite{Pasqal} to, at least in principle, attain more generality whilst still retaining the advantages of the analog computing. We will not discuss neutral atom quantum computing but instead direct the interested reader to some articles on the subject \cite{Saffman2016NeutAtom,Henriet2020NeutAtom}. In this section we will focus on the well established adiabatic quantum computation (AQC) model of computation that is used to describe computation on certain kinds of analog quantum computers such as quantum annealers. We will first discuss AQC and then move onto quantum annealing (QA) which is closely related but subtly different. There is already a large literature including many reviews on adiabatic quantum computing \cite{AQCReview} and quantum annealing \cite{QAPerspect,QAIndustryRev,QAandAQColloq,Santoro2006QA}. We will therefore restrict ourselves to a brief introduction and discussion of the key concepts. The next section will then discuss optimisation for digital (gate-based) quantum computers which is the main focus of this paper. \\

To begin an adiabatic quantum computation, a quantum system is initialized in the ground state of a simple Hamiltonian. An external field is then applied to the system and gradually changed in order to slowly transform the initial state into the ground state of a different Hamiltonian that encodes a problem of interest e.g. a combinatorial optimisation problem. The final Hamiltonian should have a ground state that corresponds to the optimal solution of the problem. The concept of AQC can be demonstrated mathematically by considering the time-dependent Hamiltonian,

\begin{equation}
    \hat{H}(t) = A(t) \hat{H}_0 + B(t) \hat{H}_1,
\label{AQC-ham}
\end{equation}

where $\hat{H}_0$ and $\hat{H}_1$ in Equation \ref{AQC-ham} are the initial and problem Hamiltonians respectively. Initially $A(t=0) = 1$ and $B(t=0)=0$, over time this changes until the reverse is true. Provided the change is gradual enough, the adiabatic theorem guarantees that the quantum state remains in the ground state throughout the computation. Formally, whether or not the time taken to implement the change (i.e. the runtime of the computation, $t_f$) is in fact gradual enough depends on the size of the energy gap between the ground and first excited states of $\hat{H}_1$, denoted $\Delta$. The smaller the gap, the more gradual the change has to be to guarantee adiabaticity. Typically $t_f$ is of order O($1/\Delta^2$), although the worst case is O($1/\Delta^3$). One of the major concerns is that this gap can become exponentially small as the problem size increases \cite{QAExpGap}. \\

Quantum annealing is very closely related to adiabatic quantum computation. Quantum annealing is essentially an adiabatic quantum computation in which the requirement of adiabaticity is relaxed somewhat. Rather than requiring the system to remain in the ground state throughout the computation, the system is allowed to enter an excited state. This relaxation is largely for practical reasons; it is easier to realise quantum annealing in hardware as opposed to a perfectly adiabatic quantum computation. Furthermore, while AQC is universal \cite{AQC-universal}, current implementations of quantum annealing are not universal. In practice, quantum annealing is actually stoquastic quantum annealing, another compromise made for the sake of practicality. Stoquastic quantum annealing is limited to stoquastic Hamiltonians, which are defined as Hamiltonians that, in a given basis, have off-diagonal elements that are real and non-positive. There is concern that stoquastic Hamiltonians are easy to simulate \cite{AQCReview} and therefore stoquastic quantum annealers may not be able to provide a quantum advantage \cite{BravyiStoq2010}. \\

One of the main difficulties that arises when developing algorithms for quantum devices (analog or digital) is how to map a problem of interest to a quantum computer. In other words, how do we translate the problem into a form that is able to be encoded in qubits and then solved by a continuous time evolution of a quantum state. Quantum annealers have largely focused on combinatorial optimisation for two reasons. The first is that such a mapping exists for many CO problems and is relatively simple. The full details are given in the section on quadratic unconstrained binary optimisation (QUBO). However, it suffices to say here that many CO problems can be mapped to a problem of the form of finding the ground state of an Ising Hamiltonian \cite{Lucas2014}. It is evident that such problems are well-suited for quantum annealers. The problem Hamiltonian of Equation \ref{AQC-ham} can simply be made an Ising Hamiltonian. The second reason is that CO problems have a huge variety of compelling real-world applications. \\

The current state of the field of quantum annealing is difficult to summarise. Despite the fact that the field is relatively young, the literature is already large and is growing quickly. It is fair to say that there is considerable debate over whether or not quantum annealers will ever achieve quantum advantage. Proponents of the QA approach are optimistic that once machines reach sufficient sizes, there is hope for an advantage in specific instances of optimisation problems. However, at the time of writing, there has been no demonstration of quantum advantage using a quantum annealer. Sceptics claim that problems such as the exponential decrease in the energy gap $\Delta$ mean that advantage will never be realised on these devices. In fact, there is a thriving cottage industry of attempting to demonstrate that quantum annealers won't be able to provide an advantage \cite{QA-pitfalls2017}. There are also many approaches that take inspiration from quantum computing to improve classical methods. For example, Fujitsu has developed a so-called quantum-inspired digital annealer. This is a purpose-built silicon chip that is designed specifically for combinatorial optimisation problems and is purportedly better at solving these problems than a classical CPU but is also easier to scale up than an actual quantum annealer \cite{FujitsuAnnealer}. In summary, the field faces considerable challenges and considerable scepticism. In spite of this, there is still a large group of scientists that are optimistic and this optimism is not entirely unfounded. Regardless of the eventual outcome, the field has generated a body of scientific work that is highly interesting and will likely continue to do so for many years to come.

\section{Optimisation on Digital Quantum Computers}

\begin{figure}
    \centering
    \includegraphics[width=0.4\textwidth]{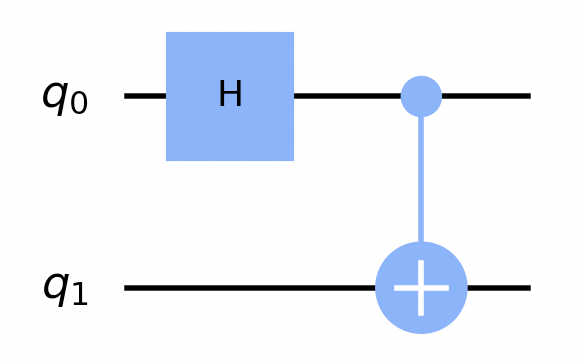}
    \caption{A simple example of a quantum circuit that prepares a Bell state, drawn using Qiskit. There are two qubits denoted $q_0$ and $q_1$, H is the Hadamard gate and the second gate is a CNOT (2-qubit) gate.}
    \label{fig:quantum-circ}
\end{figure}

Compared to analog quantum computers, digital quantum computers (DQC) are, at least in some senses, more similar to conventional classical computers. Analogous to classical logic gates, digital quantum computers implement a set of quantum logic gates (often DQC's are referred to as gate-based quantum computers). Each quantum gate corresponds to a unitary operator acting on a small number of qubits (typically 1-3 qubits) and they can be combined to build up a quantum circuit. A given quantum circuit is a representation of a computation that can be implemented on a quantum computer. A simple example of a quantum circuit to prepare a Bell state is shown in Figure \ref{fig:quantum-circ}. The quantum circuit picture is a useful abstraction largely because it is a simple and clear way of representing the evolution of a quantum state. It is often easier to understand what is happening during a quantum computation when looking at a quantum circuit compared to a long sequence of tensor products of operators. While there are many similarities between the classical and quantum gate models, there are also some key differences. For example, the fact that quantum gates correspond to unitary operators means that quantum gates must be reversible. Classical logic gates have no such restrictions. 

One of the most important concepts in the gate-based model of computing is that of universality. Formally, a set of logic gates (quantum or classical) is universal if it is able to reproduce the action of all other logic gates. For classical logic gates, the NAND gate (or the NOR gate) alone forms a universal gate set. A complication arises in the case of quantum gates because there are uncountably many possible unitary operations that could be applied to a quantum state (consider a rotation gate with an angle that may take any real value). A quantum gate set is therefore considered universal if it is able to approximate any unitary operation. The Solovay-Kitaev theorem \cite{Kitaev1997} ensures that this approximation can be to done to within a given error efficiently (the number of gates grows logarithmically as a function of the error). Broadly speaking, the benefit of universality is generality. A universal digital quantum computer can, at least in principle, be applied to solve any problem of interest. Analog quantum computers on the other hand are considerably more limited. The form of the hardware determines the exact type of problem that can be solved on the hardware. Analog quantum computers pay the price of loss of generality but hope to gain a potential advantage in certain very specific situations. It remains to be seen which approach (if either) will reach quantum advantage first. A key advantage of the digital approach to quantum computation is the potential for error correction. Encoding information digitally permits the possibility of error correction which is believed to be crucial for achieving fault-tolerant quantum computing.\\

The universality of digital quantum computers has enabled the development of quantum algorithms across a large number of application areas including simulation of many-body quantum systems, machine learning and optimisation problems. In this work, we will focus on quantum algorithms for optimisation problems, specifically combinatorial optimisation problems. Combinatorial optimisation problems are often \textit{NP}-hard to solve exactly meaning the most common approaches are heuristic methods that solve these problems approximately. There have been decades of work developing ever-improved approximate classical algorithms to solve combinatorial optimisation problems. This work has largely been driven by the fact that these problems are so widely applicable across a huge number of industries. By comparison to classical algorithms, quantum algorithms for combinatorial optimisation are still very much in their infancy. There are very few quantum algorithms for optimisation in existence and, at the time of writing, it is still very much an open question whether or not any of them will prove to be useful. \\

Despite the fact that no advantage has been established yet, there are at least theoretical grounds to think that combinatorial optimisation problems may be a suitable candidate for quantum advantage. The outcome of both classical and quantum computations is a bit string that may correspond to a solution of a given problem. However, unlike a classical computer, while a calculation is in progress on a quantum computer, the quantum state exists in a $2^{N}$ dimensional Hilbert space (where $N$ is the number of qubits). Importantly, the quantum state can be in a superposition of many states in this Hilbert space. The hope is that a well-designed quantum algorithm can extract a low-lying state (a good quality solution) from such a superposition more efficiently than classical algorithms which do not have the ability to exploit superposition. The difficulty arises because, even though a given initial superposition of states may contain the ground state (the optimal solution to the optimisation problem), if you measure the superposition, you will obtain one of the states at random. Specifically, for a superposition of $n$ states, each state will be measured with probability $1/n$. This means that superposition alone gives no natural advantage over a classical computer. Tackling this problem is the real challenge of quantum algorithm design. Fundamentally a quantum algorithm seeks to boost the probability of measuring some desired state that corresponds to a solution to a problem. In general, this is a very challenging task. Examples of algorithms that achieve this successfully include amplitude amplification \cite{Grover} which is able to take an initial superposition of states and greatly increase the probability of measuring some desired state. \\

One of the key limitations faced in the field of quantum optimisation (and more broadly in quantum computing) is the limited availability of hardware and the noisy nature of the hardware that is available. We are currently in the so-called noisy intermediate-scale quantum (NISQ) era. The combination of small numbers of qubits and noise means that only small problem sizes can be attempted on real hardware with algorithms that are low circuit depth. These small problems are typically efficiently solvable by classical means and therefore don't stand to truly benefit from quantum computers. While machines with increasing numbers of qubits are being developed at a fairly consistent pace, the issue of reducing noise is more challenging. Circuit depth is a key limitation of current hardware that makes anything but the simplest algorithms (or simplest versions of a given algorithm) impossible to implement. Without reduction in noise on the scale of several orders of magnitude, the only real hope is error correction which itself drastically increases the overhead in terms of numbers of qubits. In the NISQ era, the challenge is to develop algorithms that have some degree of resilience to noise and to identify problems that, even for small sizes, may still benefit from quantum hardware. With this in mind, many of the current approaches are hybrid quantum-classical. A suitable, difficult part of a given problem is identified and ported to a quantum computer while the part of the problem that can be efficiently solved classically remains on a conventional computer. Variational quantum algorithms fall into this category, utilising classical optimisers to find optimal parameters for quantum circuits. \\

There is a vast literature on variational quantum algorithms (VQA's) \cite{McClean2016TheoryVQA,Cerezo2021VQArev,Tilly2022VQErev}, in this work we will comment on just a few specific algorithms. VQA's are popular as they are a class of algorithms that are thought to be particularly well suited to NISQ devices. However, there is a `hidden' cost that is associated with VQA's that is not always discussed or factored into complexity theory discussions. VQA's rely on optimisation of a set of variational parameters which are typically a set of angles $\theta_i$ for single qubit rotation gates. The optimisation of these parameters is performed by a classical optimisation algorithm. The $\theta_i$ cost landscape generally has many local minima and often has flat regions (referred to as barren plateaus) that prevent further optimisation of parameters. In general, the problem of finding the global minimum is an \textit{NP}-hard problem. The cost landscape can be made increasingly favourable by overparameterising the VQA i.e. adding more $\theta_i$. However, the number of parameters (and therefore quantum gates) required to do this for some of the most common ansatzes increases exponentially making this approach impractical \cite{Larocca2021OverParam}. In this work we will discuss the specific merits (or otherwise) of a given quantum algorithm. However, when the algorithm is a variational one, it is important to bear this classical overhead in mind. \\

When using VQA's in practice, as the size of the problem and therefore the number of variational parameters increases, the classical optimisation problem can become prohibitively difficult. This issue is particularly pertinent when discussing quantum algorithms for optimisation. Often the optimisation problems that we hope to solve using a quantum computer are \textit{NP}-hard. A cynic may notice that, in the VQA paradigm, you must first solve one \textit{NP}-hard problem classically (the optimisation of the variational parameters) in order to then solve the \textit{NP}-hard optimisation problem. In other words, despite using a quantum computer, you still have to tackle an \textit{NP}-hard problem classically. This viewpoint has a degree of merit but, there are reasons to be optimistic nonetheless. Good classical optimisation algorithms exist that can efficiently provide approximate solutions. The hope is that these solutions are good enough to still yield good solutions using the VQA. \\

In the following sections, we will review the major quantum algorithms for quantum optimisation problems. We begin each section by giving an introduction to the conceptual and mathematical description of each algorithm. For the interested reader, we then detail a number of case studies. The case studies are far from exhaustive but are selected because they either show useful examples of the algorithm in action (simulated or on real hardware) or they propose interesting modifications to the core algorithm.

\subsection{Quantum Approximate Optimisation Algorithm}

The Quantum Approximate Optimisation Algorithm (QAO algorithm) was introduced by Farhi et al. \cite{Farhi2014} in 2014 and is a hybrid quantum-classical variational algorithm. Note that we refer to the quantum approximate optimisation algorithm as the QAO algorithm (rather than QAOA) in order to distinguish it more easily from the quantum alternating operator ansatz which we discuss later. In the QAO algorithm, a problem-specific cost function $C$ is mapped to a Hamiltonian $\hat{H}_C$ with an associated parameterised unitary $\hat{U}_C(\gamma) = \text{exp}(-i\gamma\hat{H}_C)$. In particular, the possible solutions to the problem are each associated with an eigenstate of the Hamiltonian such that it is diagonal in the computational basis,

\begin{equation}
\hat{H}_C|\mathbf{x}\rangle = C(\mathbf{x})|\mathbf{x}\rangle, \ \ \ \mathbf{x} \in \{0,1\}^n.
\end{equation}

The unitary $U_C(\gamma )$ corresponds to time evolution under the cost Hamiltonian $\hat{H}_C$ where the time parameter has been replaced by a variational parameter $\gamma$. In addition, a mixer Hamiltonian $\hat{H}_M$ is introduced with associated parameterised unitary $\hat{U}_M(\mu)=\text{exp}(-i\mu\hat{H}_M)$. A single block of the circuit corresponds to the application of the cost and mixer unitaries on an initial state $|\phi_0 \rangle$,

\begin{equation}
|\phi(\mu,\gamma)\rangle = \hat{U}(\mu,\gamma)|\phi_0\rangle := \hat{U}_M(\mu)\hat{U}_C(\gamma)|\phi_0\rangle.
\end{equation}

Choosing a depth hyperparameter $p \geq 1$ so that the circuit consists of $p$ blocks $\hat{U}(\mu_p, \gamma_p)$ associated with $2p$ parameters, we can now implement the overall circuit as follows,

\begin{equation}
|\phi(\mathbf{\mu}, \mathbf{\gamma})\rangle = \prod_{i=1}^p \hat{U}(\mu_i, \gamma_i)|\phi_0\rangle.
\end{equation}

The cost is then evaluated as an expectation value, 

\begin{equation}
\langle \phi(\mu, \gamma) | \hat{H}_C | \phi(\mu, \gamma)\rangle,
\end{equation}

where a classical optimiser can be used to iteratively update the parameters $\mu_i,\gamma_i$ in order to extremise the cost. The original formulation of the QAO algorithm utilised the so-called transverse-field mixing Hamiltonian which is a sum of local operators. In particular $\hat{H}_M = \sum_{i} X_i$ where $X_i$ is the Pauli $X$-operator (bit flip) acting on the $i^\text{th}$ qubit. Conceptually, the QAO algorithm can be thought of as time evolution of a quantum state combined with the power of `hopping' between states. The time evolution is generated by the cost Hamiltonian and the `hopping' is facilitated by the mixer unitary. Specifically, the mixer unitary is required to ensure that the dynamics of the system are non-trivial. If the system were evolved using only the cost Hamiltonian, the energy associated with the cost Hamiltonian (the cost) would be conserved i.e. no choice of $\gamma$ would alter the cost and therefore it wouldn't be possible to perform an optimisation. By choosing a mixing Hamiltonian that doesn't commute with the cost Hamiltonian, the cost is no longer conserved and a minimum can be found by varying $\gamma$, $\mu$. \\

Throughout the paper we will refer to a quantity called the approximation ratio that is often used to evaluate approximate methods such as the QAO algorithm. The approximation ratio $r$ is defined in terms of the maximum and minimum (optimal) cost, $C_{max}$ and $C_{min}$ respectively,

\begin{equation}
    r = \frac{C_{max}-C^{*}}{C_{max}-C_{min}}.
\label{approx-ratio}
\end{equation}

As the cost found by the QAO algorithm, $C^*$ approaches the optimal value $C_{min}$, the approximation ratio tends towards 1.

\subsubsection{Quadratic Unconstrained Binary Optimisation}

Quadratic Unconstrained Binary Optimisation (QUBO) is an \textit{NP}-hard combinatorial optimisation problem. It is primarily of interest because there are many real world problems for which embeddings into QUBO have been formulated \cite{Lucas2014}. Problems include Max-Cut, graph coloring and a number of machine learning problems to name but a few. Note that despite the utility of QUBO, it is limited by the fact that it cannot be applied to constrained CO problems. The QUBO problem is formulated by considering a binary vector of length $n$, $|x\rangle \in \mathbb{B}^n$ and an upper triangular matrix $Q \in \mathbb{R}^{n\times n}$. The elements $Q_{ij}$ are weights for each pair of indices $i,j$ in $|x\rangle$. Consider the function $f_Q : \mathbb{B}^n \rightarrow \mathbb{R}$,

\begin{equation}
    f_Q(x) = \langle x| Q |x\rangle =  \sum_{i=0}^{n}\sum_{j=i}^{n} Q_{ij} x_i x_j,
\label{qubo-definition}
\end{equation}

where $x_i$ is the i$^{\text{th}}$ element of $|x\rangle$. Solving a QUBO problem means finding a binary vector $|x^{*}\rangle$ that minimises $f_Q$, 

\begin{equation}
    |x^{*}\rangle = \text{argmin}\ f_Q(x).
\end{equation}

The difficulty of solving such a problem is readily apparent as the number of binary vectors grows exponentially ($2^n$) with the size $n$. Hard instances of QUBO are known to require exponential time to solve classically. As such, classical methods to solve these methods are typically heuristics that provide approximate solutions. The hope that is that heuristic quantum algorithms such as the QAO algorithm may be able to find better approximate solutions to QUBO problems than existing classical methods. \\

QUBO and Ising problems are closely linked and in fact a QUBO problem can be linearly transformed into an simple Ising problem (one with no external magnetic field). Ising problems are good candidates for quantum algorithms because there is a simple mapping that enables them to be executed on a quantum computer. With the goal of linking QUBO to an Ising problem and then mapping this to a quantum computer, we will rewrite the objective function in Equation \ref{qubo-definition} as a QUBO Hamiltonian,

\begin{equation}
    \hat{H}_{QUBO} = \sum_{i<j}^n Q_{ij}x_i x_j + \sum_{i=0}^n Q_{ii} x_i,
\label{qubo-hamiltonian}
\end{equation}

where the second term in Equation \ref{qubo-hamiltonian} has been simplified using the fact that, for binary variables, $x_i^2 = x_i$. The Ising Hamiltonian with no external magnetic field can be written in a similar form,

\begin{equation}
    \hat{H}_{Ising} = \sum_{i<j}^{n} J_{ij} s_i s_j + \sum_{i=0}^{n} J_{ii} s_i.
\label{ising-hamiltonian}
\end{equation}

In the Ising problem $s_i \in \{-1,1\}$ are spin variables. The similarities between Equations \ref{qubo-hamiltonian} and \ref{ising-hamiltonian} are readily apparent. A simple linear transformation of the form $s = 2x-1$ relates the two problems. Without loss of generality we therefore proceed to map the Ising problem to a quantum computer by promoting to $s_i, s_j$ to $Z_i,Z_j$,

\begin{equation}
    \hat{H}_{I} = \sum_{i<j}^{n} J_{ij} Z_i Z_j + \sum_{i=0}^{n} J_{ii} Z_i.
\label{quant-ising-ham}
\end{equation}

Note that this promotion of spin variables to Pauli $Z$ matrices is sensible because the eigenvalues of $Z$ are $\pm 1$. The Hamiltonian $\hat{H}_I$ in Equation \ref{quant-ising-ham} can now be used as the cost Hamiltonian in the QAO algorithm thereby allowing it to be applied to a large number of combinatorial optimisation problems.

\subsubsection{Case Studies}

The assessment of whether or not the QAO algorithm is a compelling candidate for quantum supremacy \cite{Farhi2019} or even advantage has been the subject of considerable research and debate. The quantum approximate optimisation algorithm is a metaheuristic, which is a procedure for generating a heuristic (approximate) algorithm. This makes it very difficult to derive any rigorous scaling behaviour outside of specific examples. It is therefore imperative that the QAO algorithm is tested as extensively as possible using simulations and, more importantly, actual hardware. In this section we will discuss a number of relevant case studies of that utilise the QAO algorithm in order to give an overview of the current state of the field. \\

Initially the QAO algorithm (depth $p=1$) was the best known approximate algorithm for MAX-3-LIN-2 \cite{Farhi2015} but a better classical algorithm was found soon after \cite{Barak2015}. Several studies have been rather pessimistic in regards to the outlook for the QAO algorithm applied to problems such as Max-Cut. For example, Guerreschi and Matsuura \cite{Guerreschi2019} explored time to solution for the QAO algorithm on the Max-Cut problem on 3-regular graphs. Here the QAO algorithm was simulated under `realistic’ conditions including noise and circuit decomposition. Their time to solution also included the cost associated with executing multiple shots for each circuit. It was demonstrated that, under these conditions, for a given time to solution the QAO algorithm (p=4,8) was only able to solve a graph 20 times smaller than the best classical solver. By extrapolation under a few assumptions, it was concluded that the QAO algorithm would require hundreds or even thousands qubits to meet the threshold for quantum advantage on realistic near-term hardware. There have been a number of other studies comparing the QAO algorithm in certain instances to classical algorithms \cite{Wurtz2021,Marwaha2022}. For example, a paper by Marwaha \cite{Marwaha2021} demonstrated that the QAO algorithm (p=2) was outperformed by local classical algorithms for Max-Cut problems on high-girth regular graphs. Specifically for all D-regular graphs with D $\ge$ 2 and girth above 5, they find 2-local classical Max-Cut algorithms outperform the QAO algorithm (p=2). \\

Despite the pessimism of the papers that have been discussed so far, there is still hope that the QAO algorithm might be able to outperform classical algorithms at higher depths than have currently been explored. Whilst Guerreschi and Matsuura's work \cite{Guerreschi2019} does not bode well for the QAO algorithm on current generation devices that, at the time of writing have at most 127 qubits (IBM Eagle), their work does suggest that the QAO algorithm might offer an advantage over classical methods once a sufficient number of qubits is reached. This claim requires an extrapolation of their data and the authors are careful to make it clear that such an extrapolation is highly uncertain due to the possibility of small-size effects in their results. An earlier paper studied the QAO algorithm for Max-Cut problems and found that the approximation ratio improved as the QAO algorithm depth, $p$ increased \cite{Crooks2018}. They compared to classical methods and found there were values of $p$ for which the QAO algorithm (simulated) was found to attain a better approximation ratio than the classical algorithm. This appears to offer some hope that the QAO algorithm will be useful at larger values of $p$. However, as Guerreschi and Matsuura point out in their paper, this study does not compare the actual time to solution of the quantum and classical methods and therefore neglects the potentially very costly process of optimising the variational parameters. Indeed Geurreschi and Matsuura take a worst case approach and extrapolate their data using an exponential due to the increasing cost of finding variational parameters as number problem size increases.\\

As we have already described, the problem of finding optimal values of the QAO algorithm variational parameters becomes increasingly difficult as the depth $p$ increases (this is a problem shared by all variational quantum algorithms). As such, there have been several papers that have attempted to address this problem by exploring improved strategies for setting parameters \cite{HadfieldFermion2018,Brandao2018,Sack2021,Galda2021,Akshay2021,Boulebnane2021InfLim}. For example, Zhou et al. \cite{Zhou2020} employed a heuristic strategy for the initialisation of the $2p$ variational parameters applied to the Max-Cut problem to reduce the number of classical optimisation runs to $O(\text{poly}(p))$. This is in comparison to random initialisation which is identified as requiring $2^{O(p)}$ optimisation runs (see the worst case exponential scaling adopted by Geurreschi and Matsuura described above). Specifically, they develop an optimisation heuristic for a $p+1$ depth circuit based on known optimisations for depth $p$. Using simulations it was shown that the optimisation curves for the values of the $p$-level variational parameters changed only slightly for $p+1$. Hence, it is possible to use results from optimisation over a $p$-depth circuit as an appropriate initialisation for the variational parameters in a $p+1$ depth circuit. \\ 

Streif and Leib \cite{Streif2020} developed a method of determining optimal parameters for the QAO algorithm that aimed to reduce or even remove the need for a classical optimisation loop. They conside applications of the QAO algorithm with constant $p$ to the problem of Max-Cut on 3-regular graphs and 2D spin glasses. They observed that the optimal parameter values tended to cluster around particular points, with decreasing variance for larger graph sizes. The optimal parameters were found to be dependent on the topological features of the problem's graph rather than the specific problem instance or system size. They use tensor network theory to develop `tree-QAOA' in which the circuit is reduced to a tensor network, associated with a tree subgraph for which contraction can be performed classically. This removes the need for calls to the quantum processing unit (QPU) when determining the variational parameters. Their numerical results for this method on Max-Cut for 3-regular graphs show results comparable to using the Adam classical stochastic gradient descent optimiser. \\

There have been a number of papers that have proposed modifications to the core QAO algorithm in order to improve performance and/or in an attempt to make the method more amenable to execution on near term devices \cite{AdaptQAOA2022,Majumdar2021,Ayanzadeh2022}. Guerreschi took a divide and conquer approach to QUBO problems using the QAO algorithm and was able to achieve an average reduction of 42\% in number of qubits required \cite{Guerreschi2021}. Zhou et al. developed an approach to solving the Max-Cut problem on devices with a small number of qubits \cite{zhou2023qaoa}. Their strategy is to partition the graph into smaller instances that are optimised with the QAO algorithm in parallel, followed by a merging procedure into a global solution also using the QAO algorithm. Their procedure, dubbed `QAOA-in-QAOA' (QAOA$^2$), potentially offers a solution to tackling large graph optimisation problems with limited qubit devices. Perhaps surprisingly, they demonstrate that their method is able to perform on par with or even outperform the classical Goemans-Williamson algorithm \cite{GWclassical1995} on graphs with 2000 vertices. Their work appears to suggest that the real power of the QAO algorithm may come when tackling large, dense graphs. However, their results are from simulations only and therefore more work is needed to determine the utility of the method on actual hardware. \\

While many studies of the QAO algorithm omit noise, there is a growing body of work examining the effects of noise on the QAO algorithm \cite{Alam2019,Alam2020,Lotshaw2022}. For example, Fran\c{c}a et al. developed a framework for studying quantum optimisation algorithms in the context of noise \cite{stilck2021limitations}. The presence of noise in a quantum computation drives the quantum state towards a maximally mixed state which corresponds to a Gibbs state with $\beta = 0$ (infinite temperature). They utilise the fact that there are polynomial time classical algorithms for sampling Gibbs states for certain values of $\beta \leq \beta_c$. As a quantum computation progresses in the presence of noise it will eventually cross over from the space of Gibbs states that cannot be efficiently sampled with a classical algorithm to the space of states that can be efficiently sampled. The circuit depth at which this happens will depend on a number of factors such as the degree of noise and number of 2 qubit gates etc. Their framework can therefore be used to estimate circuit depths for which there can be no quantum advantage on current hardware. They present such a bound for the QAO algorithm. A recent paper built on this work by examining the QAO algorithm in the context of various SWAP strategies \cite{IBMscaling2022}. Note that SWAP gates are required due to the limited connectivity of qubits in most modern QPU's. Two qubit gates can only be applied to connected qubits, if a circuit requires 2 qubit gates to be applied to non-connected qubits, then a number of SWAP gates must be employed to facilitate this. This leads to the gate count of the actual circuit that is executed on hardware being larger than the gate count of the high level circuit. This work therefore provides more realistic estimates of how noise will impact the QAO algorithm on real quantum hardware. They demonstrate that, at least for dense problems, gate fidelities would need to be significantly below fault tolerance thresholds to achieve any advantage. \\

Finally, it is important to discuss the prospects of the QAO algorithm on current or near term hardware \cite{willsch2020benchmarking,weidenfeller2022scaling}. One important study that tested the QAO algorithm on real hardware was performed by Harrigan et al. using Google's Sycamore QPU \cite{harrigan2021quantum}. They studied both planar and non-planar graph problems. The planar problems had the advantage of mapping nicely to the topology of the QPU whereas the non-planar problems required compilation using SWAP gates. They found that both planar and non-planar problem instances performed comparably during noiseless simulations. However, when tests were done on quantum hardware, the extra circuit depth incurred by compilation for non-planar problems meant that performance quickly degraded as problem size increased. By contrast, the planar graph problems were found to have performance that was roughly constant regardless of problem size. This is a useful result given that most modern QPUs have limited connectivity suggesting that only relatively simple problems can be trivially mapped to them. This casts some doubt over whether or not the QAO algorithm will be useful for more complex, interesting problems in the near-term. Further tests investigated the relationship between the QAO algorithm depth $p$ and performance. Interestingly, performance was found to consistently increase as depth increased during the noiseless simulations. However, during the tests on the real device, the extra circuit depth associated with increasing $p$ meant that noise quickly degraded performance. Despite this, the result is still a good example of empirical evidence to backup an often made claim about the QAO algorithm which is that poor performance is due to limited depth $p$ of the algorithm.\\

A different study by Streif et al. \cite{Streif2021BPSP} also provides some evidence to support the claim that increasing $p$ can lead to improved performance. They studied the binary paint shop problem (BPSP) using the QAO algorithm. BPSP is a particularly interesting test case because there is no known classical algorithm that is able to provide an approximate solution to within a constant factor of the optimal solution in polynomial time for all problem instances. They ran a mix of numerical and QPU experiments on varying problem sizes. They study the probability of achieving an approximate solution that is to within factor $\alpha$ of the optimal solution. For the hardest problem instances and at fixed $p$, they found that the QAO algorithm requires the number of samples to increase exponentially with problem size in order to maintain a constant $\alpha$ factor approximation. However, allowing $p$ to increase polynomially with system size enables a constant factor approximation factor to be maintained efficiently. This is tentative evidence to suggest the QAO algorithm can achieve better performance for BPSP than the best known classical algorithms provided $p$ is allowed to increase with problem size. However, the authors make it clear that their tests were not performed on all possible problem instances meaning no generic conclusions can be drawn.

\subsection{Constrained Optimisation and the Quantum Alternating Operator Ansatz}

One of the limitations of the QAO algorithm is that, in its original form, it can only be applied to unconstrained optimisation problems. However, many interesting combinatorial optimisation problems have constraints that must be satisfied. Hence, there is considerable motivation to move beyond unconstrained problems. One approach to account for constraints is to add penalty terms to the cost Hamiltonian that introduce large cost penalties for solutions that violate constraints \cite{Farhi2020-1,Farhi2020-2}. This is often called a soft constraint approach and it allows a constrained problem to be treated as unconstrained but significantly lowers the probability of getting infeasible (constraint violating) solutions. Solutions that violate constraints must then be pruned post facto. For unconstrained problems, the total search space of the problem is equal to the space of feasible (constraint satisfying) solutions. However, constraints reduce the size of the space of feasible solutions relative to the total search space. Soft constraint approaches may therefore become inefficient as they search parts of solution space that do not satisfy the constraints and are therefore infeasible. This is particularly problematic in cases where the infeasible solution space is much larger than the feasible solution space. \\

An alternative approach to introducing penalty terms is to construct the mixer unitary in such a way that only feasible states are explored. The quantum alternating operator ansatz (QAOA) was proposed by Hadfield et al. \cite{Hadfield2017,HadfieldFramework2022} as a generalisation of the QAO algorithm. The goal of this generalisation was, at least in part, to address the issue of optimisation problems with hard constraints. Rather than the cost and mixer unitaries taking the form of time evolution, they are able to take any form. The authors suggest certain principles that should be adhered to when designing mixer unitaries:

\begin{itemize}
    \item The feasibility of the state is preserved, that is, $\hat{\mathcal{U}}_M(\mu)$ maps feasible states to feasible states.
    \item For any two feasible basis states $|\mathbf{x}\rangle , \ |\mathbf{y}\rangle$ $\exists \ \mu : \langle \mathbf{x} |\hat{\mathcal{U}}_M(\mu) | \mathbf{y} \rangle \neq 0 $. That is, for any two feasible solutions there is always some parameterisation that allows us to evolve one state such that we measure the other with non-zero probability. In particular, it should be possible to evolve toward a more optimal feasible solution with the correct parameterisation.
\end{itemize}

The original quantum alternating operator ansatz paper presented formulations of mixer unitaries for a large number of different problems. Since then there has been considerable further exploration of this topic \cite{Ruan2020,Fuchs2021}. The hope is that, by designing problem specific mixer unitaries, it might be possible to reduce resource requirements and improve the quality of results. For example, Fuchs et al. \cite{Fuchs2022} formulated a generalisation of the XY-mixer. The family of XY-mixers \cite{Hadfield2017,Wang2020} preserve Hamming weight and are therefore useful in for problems with hard constraints. Examples include the ring mixer which utilises a Hamiltonian $H_R$,

\begin{equation}
    \hat{H}_R = \sum_{i=0}^N X_i X_{i+1} + Y_i Y_{i+1}.
\label{ring-mixer}
\end{equation}

The ring mixer applies to cases where there are $N$ qubits with periodic (ring) boundary conditions. An alternative is the complete graph mixer Hamiltonian,

\begin{equation}
    \hat{H}_K = \sum_{(i,j)\in E(K_n)} X_i X_j + Y_i Y_j
\label{complete-graph-mixer}
\end{equation}

where the sum is over all pairs of qubits on a complete graph $K_n$. The generalisation of Fuchs et al. enables the construction of mixers that preserve a feasible subspace specified by an arbitrary set of feasible basis states. The issue with their approach is that the exact way of constructing mixers scales poorly as the number of basis states increases. This is especially problematic as this number can often increase rapidly with problem size. The authors suggest the need for heuristic algorithms to approximately construct mixers which may solve this issue but, this is still an open problem.\\

Saleem et al. \cite{Saleem2023} developed a modified version of the quantum alternating operator ansatz algorithm called dynamic quantum variational ansatz (DQVA). This approach utilises a ``warm start" in which a polynomial time classical algorithm is used to find local minima that are used as inputs for the quantum algorithm. DQVA also employs dynamic switching on or off of parts of the ansatz in order to reduce resource requirements for near term devices. Their results are promising on simulators but are limited to relatively small graphs due to lack of computational resources. Note that there is a large literature on warm starting quantum optimisation algorithms that is not covered in detail here \cite{Egger2021WarmStart,Tate2023WarmSDP,Okada2022Systematic}.

\subsubsection{Physics inspired approaches}

One of the more interesting strategies for designing mixers in the quantum alternating operator ansatz framework is the method of appealing to physics for inspiration. In this section we will discuss several papers that take this approach to design mixers for both physics and non-physics applications. For example, Kremenetski et al. \cite{Kremenetski2021} applied QAOA to quantum chemistry problems. The authors utilise the full second-quantized electronic Hamiltonian as their cost function. Their mixer unitary is time evolution generated by the Hartree-Fock (HF) Hamiltonian. This mixer naturally preserves feasibility as states can only evolve into other chemically allowed states. The method is tested on P$_2$, CO$_2$, Cl$_2$ and CH$_2$ using classical simulations. The paper represents a nice proof of concept of a physics inspired mixer for a physics application of QAOA. \\

Another recent paper applied the quantum alternating operator ansatz to the problem of protein folding on a lattice. Building on previous quantum annealing work, \cite{Babej2018} Fingerhuth et al. developed a novel encoding for a lattice protein folding problem on gate-based quantum computers \cite{Fingerhuth2018}. The method utilises ``one-hot" encoding to encode turns on a lattice in order to simulate protein folding. They also develop several problem specific mixer Hamiltonians that encode constraints that, for example, ensure the protein does not fold back on itself. The authors find that their results improve significantly when they initialise the system in a superposition over feasible states rather than a superposition over all states.  The authors demonstrated a proof of concept of their method on a Rigetti 19Q-Acorn QPU. However, hardware constraints meant that they had to split the problem into parts and had to initialise the state as a uniform superposition. They achieved a 6.05\% probability of reaching the ground state in their test on the real quantum device. \\

There has been some work exploring symmetries in the context of QAOA \cite{Shaydulin2021}. Recent work by Zhang et al. exploits the fact that symmetries and conserved quantities are intimately linked \cite{Zhang2021}. In this case, the authors exploit gauge invariance in quantum electrodynamics (QED) to formulate a mixer Hamiltonian that naturally conserves flow in the context of network flow problems. Network flow problems satisfy the constraint that flow must originate and terminate only at sources and sinks respectively. A consequence of this constraint is conservation of flow i.e. the amount flowing into a given node must equal the amount flowing out of a given node (unless it is a source or sink). The authors draw an analogy between these properties of network flow problems and lattice QED. The analogy is justified by noticing that, when Gauss' law $\nabla\cdot \vec{E} = \rho$ is written in discrete form, it is of the same form as Equation \ref{max-flow-1}. The authors therefore equate positive and negative charges to sources and sinks respectively. Electric charge is equivalent to the amount of goods and the electric field $\vec{E}$ represents a flow. \\

In order to derive their mixer Hamiltonian, Zhang et al. write down a minimal gauge invariant lattice QED Hamiltonian. Making use of the analogy outlined above, the authors propose this as a QED-mixer Hamiltonian for network flow problems. The gauge symmetry of the Hamiltonian means that flow is naturally conserved without the need for extra penalty terms. Despite satisfying the constraint of flow conservation, the QED-mixer Hamiltonian can generate solutions that contain isolated closed loops of flow that are unconnected to any source or sink and are therefore not feasible solutions. The authors therefore propose a modified, restricted QED-mixer that solves this problem at the cost of increased circuit complexity.\\

Zhang et al. show that their mixer performs well compared to the standard X-mixer for a number of relatively simple test cases. Specifically they study single source shortest path (SSSP) problems and edge-disjoint path (EDP) problems on undirected graphs. All test cases were performed using numerical simulations rather than actual quantum devices. Despite the fact that the test cases are relatively limited in scope, the paper clearly demonstrates that potential of developing physics-inspired mixers. Considerable research effort relating to QAOA involves the development of improved mixers and physics inspired approaches like the ones discussed in this section offer a clear, rigorous path to improvements.

\subsubsection{Case studies of Constrained Optimisation}

Case studies of QAOA with hard constraints on actual quantum devices have been somewhat rare due to hardware limitations both in terms of number and fidelity of qubits. However, as hardware continues to improve, QAOA is increasingly able to be tested. For example, Niroula et al. \cite{Niroula2022} recently performed an insightful study comparing unconstrained (penalty term) QAOA, XY-QAOA and the layer variational quantum eigensolver (L-VQE) algorithm. They studied the problem of extractive summarisation (ES) which is an example of a constrained optimisation problem that has real world application. Tests were performed using noiseless and noisy simulators as well as Honeywell's H1-1 20 qubit trapped ion quantum device. In their tests, both L-VQE and QAOA account for constraints by including a penalty term whereas in XY-QAOA the constraints are encoded directly into the mixer. The XY-QAOA method also requires preparation of an initial state that is in a superposition over all feasible states. 14 and 20 qubit experiments were performed with 10 repetitions for all methods apart from XY-QAOA which was only repeated 3 times due to high circuit depth. Their results contain several key insights into how best to treat constraints and the impact of noise. \\

Across all platforms (noiseless/noisy simulator and QPU) the penalty term QAOA had very poor probability of generating feasible solutions (worse than random in the 14 qubit tests). The authors show that this poor performance arises because introducing a penalty term leads to a trade-off between terms in the Hamiltonian. L-VQE doesn't suffer from the same issue in these test cases because it is expressive enough to be able to solve the problems exactly. However, this won't be the case in general for L-VQE when applied to larger problem instances that become more difficult to optimise as the number of parameters increases. These tests also demonstrate that XY-QAOA is highly sensitive to noise. On the noiseless simulator, XY-QAOA delivers feasible solutions with 100\% probability but on the noisy simulator and quantum device, the probability drops to less than 40\%. The authors suggest that this is likely due to the increased circuit depth and number of 2 qubit gates required for XY-QAOA relative to both the other methods. In particular, the state preparation step for XY-QAOA is non-trivial and already introduces noise. \\

These results appear to support the idea that, at least for QAOA, it is crucial to setup the mixer unitary in such a way that it naturally encode constraints. However, the increased cost in terms of circuit depth and number of 2 qubit gates associated with this approach means that such implementations are more susceptible to noise and are therefore less likely to be useful on current generation devices. The comparison between L-VQE and XY-QAOA in this instance is inconclusive as, for the small problems studied, the two methods perform comparably.\\

Baker et al. \cite{Baker2022} were the first to test QAOA with hard constraints combined with initial preparation of a superposition of feasible states (Dicke state) \cite{Aktar2022} on a QPU. Note that the aforementioned protein folding study \cite{Fingerhuth2018} came first and utilised hard constraints but the initial state preparation was not performed on the QPU. Baker et al. solved mean-variance portfolio optimisation (MVPO) problems using QPU's from multiple hardware providers: Rigetti (Aspen 10), IonQ (11 qubit device) and IBM (multiple devices). Their tests provide a direct comparison between two kinds of hardware: superconducting and trapped ion, each of which have different qubit connectivities. The test cases varied both QAOA depth $p$ and the problem size. The methods tested were soft constraint QAOA (penalty term) as well as two hard constraint mixers: ring and complete graph, with random and Dicke state initialisation respectively. \\

The tests performed were comprehensive and worth reading in full detail. Here we will summarise some of the key results. In a large number of cases QAOA outperforms random although, as $p$ and problem size increase, this is not necessarily always the case. QAOA with soft constraints performs best across the QPU tests. However, the authors state that the total number of gates required for the hard constraint methods was roughly an order of magnitude larger than for the soft constraint circuits. They therefore attribute the poor performance of hard constraints to noise. This is backed up by the fact that the hard constraint approach with random state initialisation, which requires fewer gates, out performs the Dicke state approach. This further confirms the findings of Niroula et al. that, while hard constraint QAOA may be preferable, the increased circuit depth makes the approach less suitable for current devices. The authors also find a concerning variability in result quality across repetitions on all QPU's (regardless of hardware type) that far exceeds shot-based stochastic variability. \\

Finally, a paper by Wang et al. deserves a brief discussion \cite{Wang2020}. They performed numerical simulations of penalty term and XY-mixer QAOA for graph colouring problems. Consistent with the findings already discussed, they discover that penalty term QAOA has a poor approximation ratio compared to the XY-mixer. In their tests, they also find that the complete graph mixer outperforms the ring mixer. Interestingly they demonstrate that increasing QAOA circuit depth consistently improves approximation ratio. However, as we have already seen, the effects of noise can't be ignored for QAOA. These tests therefore need to be considered in the full context of the studies that utilised real hardware.

\subsection{Quantum Semidefinite Programming}

Semidefinite programming (SDP) is a field of convex optimisation that is interesting because a number of practical problems (e.g. combinatorial optimisation) are able to be modelled or approximated using SDP. In this section we will outline the basic formulation of an SDP problem and then discuss recent work that has been done to develop quantum algorithms for semidefinite programming. Given matrices $(A_1,..., A_m, C) \in \mathbb{S}^n$ where $\mathbb{S}^n$ is the space of symmetric $n\times n$ matrices and real numbers $(b_1, ... , b_m)$ a general SDP problem takes the form,

\begin{equation}
    \min_{X \in \mathbb{S}^n} \langle C, X\rangle_{\mathbb{S}^n},
\label{SDP-1}
\end{equation}

subject to the constraints,

\begin{equation}
    \langle A_i, X \rangle_{\mathbb{S}^n} \leq b_i.
\label{SDP-2}
\end{equation}

Where $X$ is a positive semidefinite matrix, denoted $X \succeq 0$. Positive semidefinite matrices are self-adjoint matrices that have no negative eigenvalues.
Stated in words, the SDP problem requires finding a matrix $X$ that minimises the cost which is given by the inner product in Equation \ref{SDP-1}, while satisfying the constraints (Equation \ref{SDP-2}). There are many classical algorithms to solve SDP problems \cite{LeeSDP2015,AroraSDP2005,AroraSDP2012}. We have already mentioned one example when discussing QAOA: the GW algorithm for Max-Cut \cite{GWclassical1995}. It is important to note that SDP applies only to convex optimisation problems. Convex problems have a single minimum which, in most cases, makes them easier to solve than non-convex problems that typically have many local minima. While many problems that are of real world interest are non-convex (e.g. optimising variational parameters for VQA's), SDP is useful nonetheless. Often SDP can be used to approximately solve problems that are not necessarily convex by performing a so-called `convex relaxation'. This is a relaxation of the original problem in order to make it convex. GW is an example of an algorithm that makes use of this procedure.\\

To date there have been only a relatively small number of papers that propose quantum algorithms for semidefinite programming \cite{Chakrabarti2020,vanApeldoorn2020,Kerenidis2020,BhartiNISQ2022,Huang2023}. Semidefinite programming as a field is relatively young and quantum semidefinite programming is even more so. The considerable success of classical SDP algorithms has motivated the exploration of quantum SDP algorithms. In the remainder of this section we will briefly discuss a small number of these works. Brand\~{a}o and Svore proposed a quantum SDP algorithm in 2017 \cite{BrandaoQSDP2017} that utilises a combination of Gibbs state preparation and amplitude amplification. They utilise the fact that it is possible to solve an SDP problem by preparing Gibbs states of Hamiltonians given by linear combinations of the input matrices $A_i,$ and $C$. They then utilise amplitude amplification to reduce the cost of preparing the Gibbs states. Note that their paper presents just the algorithm without any simulations. Therefore, despite the authors giving bounds on properties such as run time, we don't know whether or not the algorithm is practically useful. \\

Bharti et al. developed a quantum SDP solver specifically targeting NISQ hardware \cite{BhartiNISQ2022}. They achieve this by developing a variational quantum algorithm. One of the most interesting aspects of the paper is their approach to optimising the variational parameters. The cost landscape is typically highly non-convex and finding the global minimum is an \textit{NP}-hard problem. Classical optimisation of variational parameters is therefore often plagued by issues such as local minima and barren plateaus. Bharti et al. use a classical SDP solver to optimise their variational parameters. This setup limits the dimension of the classical SDP problem to that of the variational ansatz rather than the actual quantum SDP problem that is being solved. However, SDP can only be used for convex optimisation problems and optimising VQA's is known to be non-convex in general. As such, Bharti et al. formulate the problem in terms of density matrices. They consider a hybrid density matrix, $\rho = \sum_{i} \beta_i |\psi_i\rangle\langle\psi_i|$ where $\beta_i$ are the parameters optimised by the classical SDP. It is shown in the paper that this is a convex optimisation problem and as such a classical SDP solver can be used in order to solve the problem efficiently.

\section{Discussion \& Analysis}

In this paper we have explored some of the most popular quantum optimisation techniques currently being considered, particularly with a view to deployment on near term (noisy intermediate scale) quantum devices. In both classical and quantum optimisation, the techniques used are most commonly heuristic methods that provide approximate solutions. Applications of quantum optimisation are still limited to what could be considered toy problems, and it remains to be seen how these approaches may be translated into practical industrial application of quantum advantage. Indeed, as much of the literature shows, there are still strides to be made when it comes to realising quantum advantage in a narrower context. Nonetheless it is possible to identify some of the potential advantages and shortfalls inherent in these techniques, which we hope will guide practitioners in the field toward the development of new effective methods for yielding quantum advantage. \\

In this work we have discussed two of the main quantum hardware paradigms that currently exist. Quantum annealers represent an analog approach, deploying continuous time evolution to solve optimisation problems that can be appropriately translated into a well-behaved quantum mechanical problem. However, the use cases remain constrained, and existing quantum annealers are not able to implement universal quantum computation. Nonetheless, the application of quantum annealers to discrete optimisation problems has captured a great deal of interest in part owing to the availability of D-Wave Systems' commercial platform. Typically quantum annealing is considered in the context of adiabatic quantum computation (AQC), where the system is evolved slowly between the ground states of two Hamiltonians, with the final state representing the solution of the optimisation problem. The adiabatic theorem guarantees that the system remains in the ground state as long as the schedule is sufficiently slow. However, in many practical cases this restriction is unrealistic. Small gaps between the ground state and first excited state energy can lead to long annealing times or error-prone computation which may negate quantum advantage. However, proposals have been made to allow annealing to take advantage of small jumps in energy, especially early in the annealing process \cite{crosson2021prospects}. This `diabatic' annealing pathway shows some promise due to its more general regime and resistance to classical simulation. \\

The other hardware paradigm, and the main focus of this paper, is the ``digital" or gate-based quantum computer. In contrast to annealers, digital quantum computers offer universal quantum computation. While digital quantum computers are distinct from their analog counterparts in many ways, there are also similarities. For example, the quantum state of a digital quantum computer still undergoes a continuous evolution through time from an initial to a final state making it analog in a sense. Despite this, certain guarantees on the approximation of gates and the separability of their operations make the analogy to digital circuits a useful abstraction for modelling quantum computation. To date there have been many quantum algorithms proposed that utilise the quantum circuit model. Many promise to enhance classical methods or to provide quantum advantage but in practice they are hampered by the limitations of current hardware.\\

The quantum approximate optimisation algorithm and the quantum alternating operator ansatz framework represent the vast majority of the field of quantum optimisation on digital quantum computers. Since their inception, these methods have dominated the field of quantum optimisation. We have discussed a wide variety of papers that implement variations of the quantum approximate optimisation algorithm as well many papers operating in the quantum alternating operator ansatz framework. Despite this, we have barely scratched the surface of this literature. Across the papers that we have reviewed in this work, there are some common themes. In many cases hardware limitations are blamed for poor performance of QAOA or for the lack of testing of QAOA on hardware. The noise of qubits limits circuit depth and therefore the QAOA depth $p$. Furthermore, the small number of qubits available limits the size of problems that can be studied to regimes that, for all optimisation problems of real world interest, are efficiently solvable classically. These are valid concerns and there is evidence to suggest that improving fidelity and number of qubits will improve the utility of QAOA. However, there are several key studies we have mentioned that suggest improvements in fidelity and qubit counts will have to be substantial in order for QAOA to yield any advantage \cite{Guerreschi2019,IBMscaling2022}. If these studies are correct in their assessments, we will require, at best, several further generations of quantum computers to be developed or at worst fault tolerant devices to be developed before we can expect advantage. Furthermore, even if qubit counts increase and fidelities improve, the corresponding increase in number of variational parameters will start to cause serious problems for the classical optimisation part of the algorithm. Efforts have been made to reduce the overhead in the classical optimisation loop including heuristics to `pre-optimise' the parameters for circuit based on values found in a shallower circuit \cite{Zhou2020}, or by considering the topology of the optimisation circuit as a tensor network \cite{Streif2020}. Despite this, there is no guarantee that we will be able to find good variational parameters efficiently as problem sizes and depths grow. It is therefore far from certain that simply waiting for better hardware to be available will naturally lead to advantage at some point in the future. \\

Another common theme in the papers we have discussed is the use of noiseless simulators when testing implementations of QAOA. While many of these studies are interesting proofs of concept, the lack of noise makes them highly unrealistic if the goal is to assess the prospect of quantum advantage. We have discussed a small number of papers that studied noise in a realistic manner or even ran on real hardware. In general, the conclusions are that going beyond low depth implementations of the basic quantum approximate optimisation algorithm is essentially infeasible on current hardware due to the impact of noise \cite{Niroula2022,Baker2022,harrigan2021quantum}. This is problematic in the context of constrained problems because the evidence we have reviewed appears to suggest that hard constraints are in fact necessary to get good quality results. However, the hard constraint approach increases circuit depth to the point that executing on noisy hardware is no longer possible. For the foreseeable future quantum computers will be noisy and it is therefore crucial for any algorithm that hopes to obtain near-term advantage to be resilient in the presence of noise. The research that has been done to date highlights how crucial it is to test on real hardware or, at minimum, with realistic noise models. \\

Although our assessment of QAOA so far has focused on the challenges and reasons to be pessimistic, we believe there are also reasons to be optimistic. There is some evidence, both empirical and numerical, that suggests the performance of the QAO algorithm improves as $p$ increases \cite{Streif2021BPSP,harrigan2021quantum}. We have also reviewed a considerable number of papers that operate in the quantum alternating operator ansatz framework, designing problem-specific mixers. It is possible that there exist certain highly specific optimisation problems for which an effective and efficient problem-specific mixer can be designed. This may be an avenue to achieving advantage sooner, although by nature any such advantage would be limited in scope. Another practical avenue for near-term advantage is the utilisation of multiple QPU's in parallel. We have discussed some papers that employ strategies of breaking up problems into smaller parts \cite{Guerreschi2021,zhou2023qaoa}. If approaches like this are able to be parallelised across multiple QPU's then it is possible that some of the hardware limitations of current generation devices can be circumvented sooner than expected. \\

Although QAOA is by far the most popular method for quantum optimisation, there are a number of different methods that exist but are less mature. We have briefly discuss quantum semidefinite programming which is one such example. Quantum SDP can be applied to convex optimisation problems which excludes a number of problems of interest that are non convex. However, quantum SDP is still a promising and growing area of research. The hope is that the success of classical SDP in recent decades can be replicated in the quantum realm. Beyond the scope of what we have discussed here, other optimisation techniques have been explored in the literature including the use of branch \& bound algorithms \cite{Montanaro2020}. Other examples include a recently proposed technique of logarithmic encoding that considerably reduces the resource overhead in solving many \textit{NP}-hard graph problems \cite{Rancic,Chatterjee2023}. There is also a small but growing literature on Quantum Metropolis sampling, first proposed by Temme et al. \cite{Temme2011QMS} The algorithm is inspired by the classical metropolis algorithm and it can be applied to optimisation problems among other things. Since the original method was proposed, there have been a number of developments \cite{Yung2012QQMS,Shtanko2021Preparing}. Chen et al. provide a thorough overview of the developments to date in their recent paper \cite{Chen2023QTSP}.\\

Overall optimisation is a promising application area for quantum computing, not least because the potential real world gain is huge if quantum advantage were to be achieved. There are reasons to be optimistic that quantum computers will be able to efficiently solve optimisation problems, at least principle. However, in practice there has been no realisation of this potential advantage and there are many reasons to be sceptical that such advantage is even close to being achieved. In order to advance the field we require improved, noise resilient algorithms as well as techniques that utilise hardware in more creative ways. In order to achieve this, there is a desperate need for thorough benchmarking of existing methods using noisy simulations and real hardware in order to better understand the impact of noise and performance on NISQ hardware. Such benchmarks would be invaluable in guiding future research efforts in the field. \\

\begin{acknowledgments}
This work was supported by the Hartree National Centre for Digital Innovation, a UK Government-funded collaboration between STFC and IBM.
\end{acknowledgments}




\bibliography{quantumoptbibliography}

\end{document}